# The Impact of an AirBnB Host's Listing Description 'Sentiment' and Length On Occupancy Rates

RICHARD DIEHL MARTINEZ, ANTHONY CARRINGTON, TIFFANY KUO, LENA TARHUNI, NOUR ADEL ZAKI ABDEL-MOTAAL

## Abstract

There has been significant literature regarding the way product review sentiment affects brand loyalty. Intrigued by how natural language influences consumer choice, we were motivated to examine whether an AirBnB host's occupancy rate (how often their listing is booked out of the days they indicated their listing was available) can be determined by the perceived sentiment and length of their description summary. Our main goal, more generally, was to determine which features, including (but not limited to) sentiment and description length, most influence a host's occupancy rate. We define sentiment score through a natural language algorithm process, based on the AFINN dictionary. Using AirBnB data on New York City, our hypothesis is that higher sentiment scores (more positive descriptions) and longer summary length lead to higher occupancy rates. Our results show that while longer summary length may positively influence occupancy rates, more positive summary descriptions have no effect. Instead, we find that other factors such as number of reviews and number of amenities, in addition to summary length, are better indicators of occupancy rate.

## I. Introduction

***About AirBnB***

AirBnB, founded in 2008, is a unique lodging and hospitality service. It is an online marketplace allows for travelers to find accommodations in homes, apartments, hostels, and even castles for days, weeks, or months on end. Furthermore, individuals with extra space in their homes are able to monetize on the lodging needs of these travelers. With over 3 million listings in over 191 countries, AirBnB has satisfied the lodging needs of over 150 million guests.

***Listings & Occupancy Rate***

In order to rent out spaces, hosts must advertise their lodging on the AirBnB platform. A posting consists of a listing title, an "about this listing" section, features of the space (including rules), amenities, prices and cancellation policy, house rules, and safety features. Additionally, there is a description section dedicated to describing sleeping arrangements, the space, interactions with guests, the neighborhood and methods of getting around. While this is the bulk of the information provided by the host, potential guests are also able to view reviews from previous guests. We will analyze the causal relationship between a host's listing description length and 'sentiment', and the likelihood of their listing being booked more often (occupancy rate).

***Research Question***

Which features most influence a host's occupancy rate and can be recommended for hosts to improve to increase their bookings? More specifically, does the perceived sentiment and length of an AirBnB host's summary description causally affect their occupancy rates in New York City? Sentiment refers to how positive or friendly the text is and is gauged via the AFINN dictionary. Using this AirBnB dataset[1] we are interested in examining how the length and sentiment of an AirBnB host's description summary affects the occupancy rate of their listing. Our hypothesis is that longer summary lengths and higher sentiment scores (indicating more positive descriptions) will positively influence a host's occupancy rate.

If a positive correlation were to exist, we would expect longer messages to include more 'positive' words, and thus, have a more positive sentiment, while shorter messages are more likely to be brief and may not give off a positive sentiment to potential tenants. However, it is important to consider that while a host's introduction may be longer in length, this does not necessarily mean they are as friendly as one may also assume. We may end up finding that the length of the message is very weakly (positively) correlated to sentiment. The information we will gather is important and relevant because AirBnB hosts could use it to increase their chances of their listings being booked.

## II. Literature Review

A wealth of sentiment-analysis studies have been conducted to assess the impact of the individual customer's opinions on prospective customers. Sentiment analysis refers to the examination of natural language used in evaluations and appraisals. This is expressed via attitudes and emotions towards factors such as products and services (Liu 1). Generally, this analysis is conducted through creating a polarity in between sentiments by classifying them as positive or negative. Prior to 2000, the research horizon of sentiment analysis particularly for commercial purposes was sparse (Liu 2). Through realizing the versatility for applications of linguistics and natural language processing and the expansion of consumer presence on the web (post tech-boom), the indications of consumer reviews were realized as a tool to allow businesses to understand their consumers and what to improve upon in future product or service development. Much of the current research horizon utilizes customers' reviews as a means of identifying product feature opinion and the impact on sales.

First of all, research conducted by Suryadi and Kim, aimed to identify which product features were dependent on sentiment based off of the reviews. They performed a correlational analysis of the content of reviews related to relevant product features and sales rank data. They tagged different parts-of-speech to be able to define the relevant product feature and opinion, combined the data with numerical ratings and utilized it to explain sales rank. Overall, they found that this analysis implicitly reflects consumer motivations in purchasing specific products. This relates to an advanced front of sentiment analysis in the ability to determine what features are being assessed based off of reviews and ultimately determining consumer thoughts. While Airbnb has an immense amount of reviews, this could lead to a better understanding for formulating listing descriptions and what factors such as amenities should be emphasized.

A more sentiment-focused word-use approach to Suryadi and Kim's research was conducted by Reinhold Decker and Michael Trusov. In this research, Decker and Trusov aimed to decipher aggregated consumer preferences through a sentiment analysis of 20,000 consumer reviews of mobile phones in Germany. The first step to this research was to establish words in the pros or cons category. To do so, the process included the following[2]:

1. Partition words and phrases of reviews into pros and cons
2. Eliminate words that do not provide any explicit or implicit information of product features

[1] Appendix 1: AirBnB Dataset (http://insideairbnb.com/get-the-data.html)
[2] Appendix 3: Chart of Frequency of Features

3. Aggregate redundant words and phrases that include a feature description as a pro or a con. (Ex: “Amazing speakerphone” and “Excellent speakerphone” aggregate as pros under the label “speakerphone”)
4. Alter implicit descriptions into explicit features (Ex: “Expensive” to “Price”)
5. Merge synonyms of a feature to a single word for the feature
6. Elimination of low frequency mentioned features
7. Transform pros and cons to binary listings

Researchers used three different models to assess econometric preference with the main contributing model being heterogenous[3] with a discrete distribution. Using a negative binomial regression, researchers found strong correlations between consumer opinion reviews and consumer preferences. This regression produced estimates of parameters indicating “relative effect of functional attributes and brand names on product evaluations and purchase decisions.”(Decker and Trusov 304). When applying their findings, they found that their hit rate was 51.5%, nearly twice as much as hit rate by chance (Decker and Trusov 304).

Overall, this study of a user product, which is heavily used and reliant on consumer interaction, indicated important factors for product development. In the case of AirBnB, most hosts have reviews, however, another basis for assessing how consumers would decide on a particular host’s accommodation is through another dimension of gauging differing sentiments from different listings and the hosts they will be present with and/or interact with during their stay. Therefore, while much research regarding the impact of reviews has been conducted since 2000, it is this additional factor of a service being proposed and offered through the personal attributes of the service-provider and the sentiment that the potential tenant may have.

Chevalier and Mayzlin (2006) conducted research in a similar manner on online marketplace platforms for Barnes and Noble and Amazon. Online marketplaces have become a main destination for consumers and AirBnb has furthered that with altering the landscape for booking accommodations via the web with a disruptive technology-esque approach. In this specific study, the reviews of customers are assessed to see if they differ in impact between the sites, and if the reviews truly do impact sales(Chevalier and Mayzlin 345). Researchers looked at multiple coefficients including price and rankings as indicators of effect of sales. Ultimately their as-if experiment resulted in them finding that even between marketplaces that offer the same products, the reviews correlated with sales and that negative reviews had a greater impact than positive reviews on the particular site the review was written (Chevalier and Mayzlin 353).

This research into the effects of “word of mouth” on current online marketplaces and consumer behavior indicate that there may be a connection amongst the verbiage that individuals use to communicate with online consumers overall and the impact on sales, or occupancy rate in the case of an AirBnb host. Further, while all listings differ, the similarities in offerings of homes within zip codes, or perhaps accommodations with similar amenities could result in differing occupancy rates. It will be important to see how these different offerings and their intertwining with listing description may ultimately affect their occupancy rate, or in the context of the study, the sales rate.

After having assessed this literature, it is clear sentiment expressed via web reviews simulates a person-to-person interaction that strongly correlates with business outcomes. In our research, we are able to fuse these ideas with the more distinct business model of AirBnb; where the unique product description of each listing is designed to simulate that same person-to-person interaction of a review and ultimately may affect occupancy rate.

---

[3] Heterogeneity accounts for the multiple and diverse uses of products which is common in consumer practice

# III. Data Methodology & Analysis

***Data Cleaning***

The New York City (NYC) AirBnb listing dataset included 95 variables for each listing.[4] We artificially selected features that we believed would influence occupancy rate, resulting in a new dataframe listing of the feature columns, "property type, number of people accommodated, number of bathrooms, number of bedrooms, number of beds, bed type, price, price per occupant, number of reviews, zip code, length of summary description, number of amenities, length of space description, and overall rating. We calculated a 'sentiment score' (process described below) for the summary and space descriptions, and added two columns for 'sentiment score of summary description' and 'sentiment score of space description, giving a total number of 15 features. We replaced N/As with 0s for summary/space description lengths, and replaced N/A overall ratings with the mean rating of 4.62.[5]

With a dataset detailing occupancy rate for AirBnb listings in NYC purchased from AirDNA, a company that sells AirBnB data, we conducted an inner join with our clean listing data. Using *host_id* to match, we added a column for occupancy rate, and listings without a recorded occupancy rate.

***Sentiment Scores***

We calculated sentiment scores using the AFINN-111 dictionary. AFINN is a list of 2477 English words that have been rated by sentiment (from a lowest possible sentiment of -5 to a highest sentiment of +5). This database was created by Finn Arup Nielsen in 2011, and published by the Technical University of Denmark. Throughout our analysis of sentiment scores, we naively assume that the sentiment values listed in the AFINN dictionary are applicable to the AirBnb description data. In order to calculate the positive or negative sentiment of each listing's description, we counted the number of times that each word in the AFINN databased occurred in each of the listings, and multiplied these counts by the sentiment scores provided by AFINN.[6] Although simple, this approach clearly highlighted the difference between listings whose descriptions conveyed a much stronger positive sentiment than others. For instance, below we showcase a description with a sentiment score of -7, and one with +35. The difference is immediately apparent.

**Sentiment Score of 35**: *Welcome to Harlem USA! Harlem heights is the new point of transformation in NYC and my apartment has a great location! close to many trains and close to many businesses as well as parks and tourist attractions. it's very cool uptown and the private room in the apartment is a great size.*

**Sentiment Score of -7:** *prospective guests must provide full name(s) to host guests occupying premises agree to indemnify hosts against any injury, illness, damage or loss during residence and not take legal action for any of these.*

Despite the preliminary success of our sentiment calculation method, we acknowledge that we fail to incorporate an analysis of the context and grammatical structures that go into determining the positive sentiment of a description. As a result, our rough estimation for the sentiment score of a listing is likely to suffer from biases and will fail to capture the nuances of the language used in the description.

[4] Appendix 1: Original AirBnB Dataset
[5] Appendix 2: Data Cleaning
[6] Appendix 4: Sentiment Score Methodology

Modern natural language processing algorithms may be used in future research in order to find more accurate approximations for the sentiment of a description.

***Frequencies and Correlations***

We began our data exploration by visualizing different relations amongst our data to derive some intuition about which features in our data are correlated with occupancy rate. The first step in this process was to observe both the frequency of sentiment scores across all of the listings data, as well as the frequency of the summary lengths. Plotting these two distributions, gave the following visualizations:

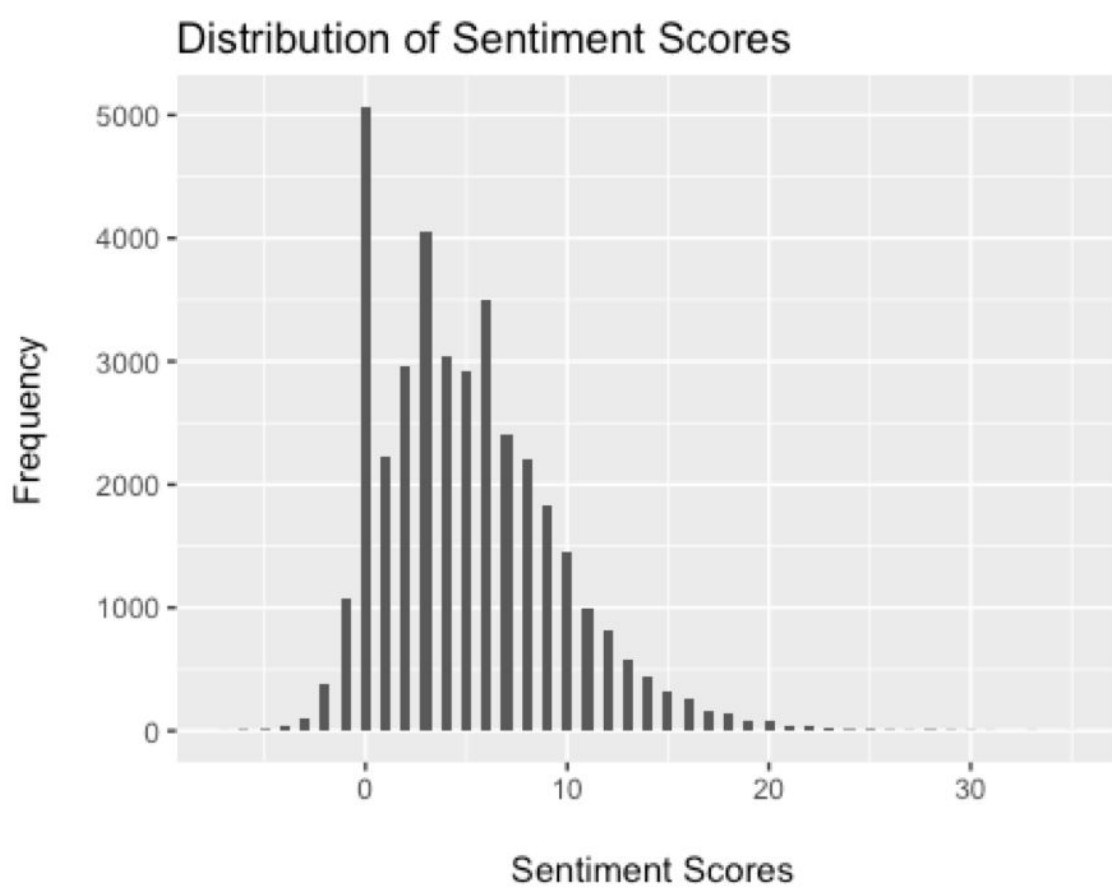

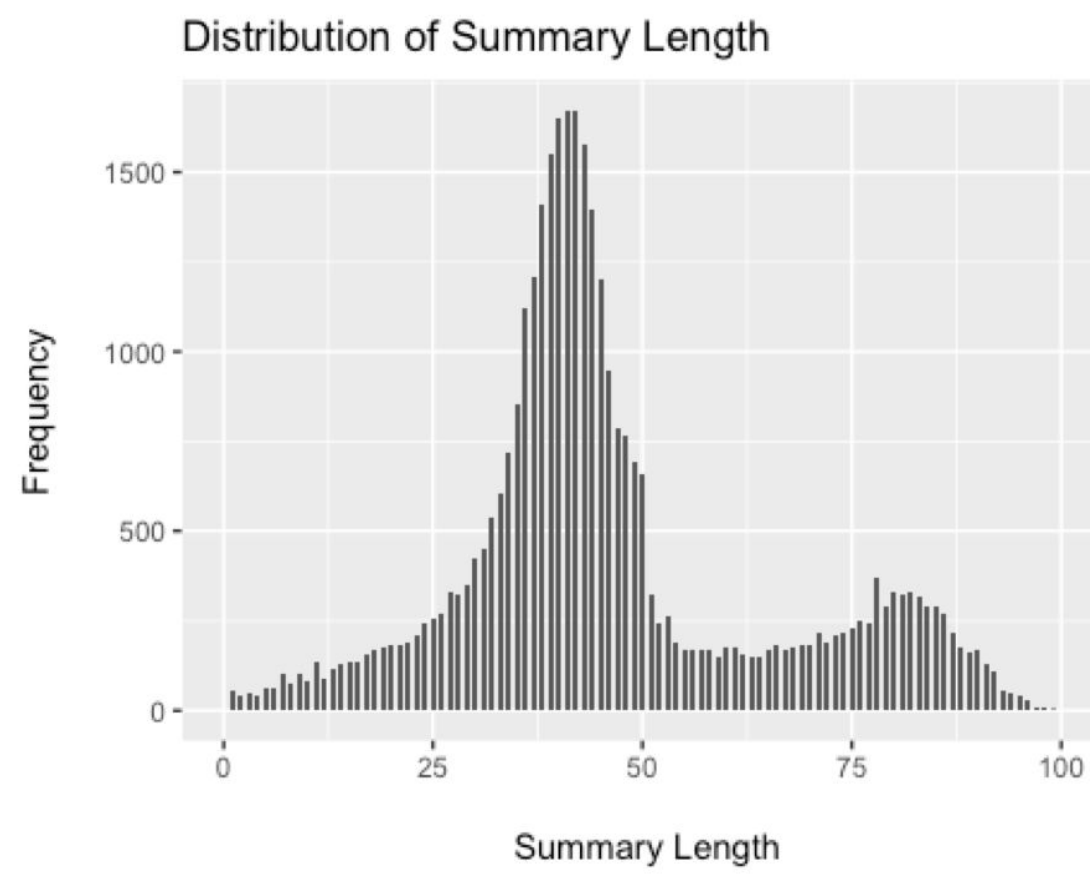

These graphs illustrated several points. For one, the majority of the data is clustered tightly around the mean of the distributions (Mean Sentiment Score: 5, Mean Summary Length 44.4 words). We also note that the overall distribution of the data point is approximately bell-shaped, with a few key outliers both at the upper and lower ends. Somewhat concerning is the fact that although we observe a large peak in the number of listings' descriptions with 0 sentiment score, we note that only very few of these descriptions have a length of 0. We can thus infer that for a large proportion of our listings, our sentiment score calculation algorithm assigned these descriptions a score of 0, which seems unrealistic and provides some indication of the inaccuracy of our sentiment calculation methods.

We continued our data exploration by plotting occupancy rate versus sentiment score, expecting (according to our original hypothesis) to observe a strong positive trend. Instead, we found the following:

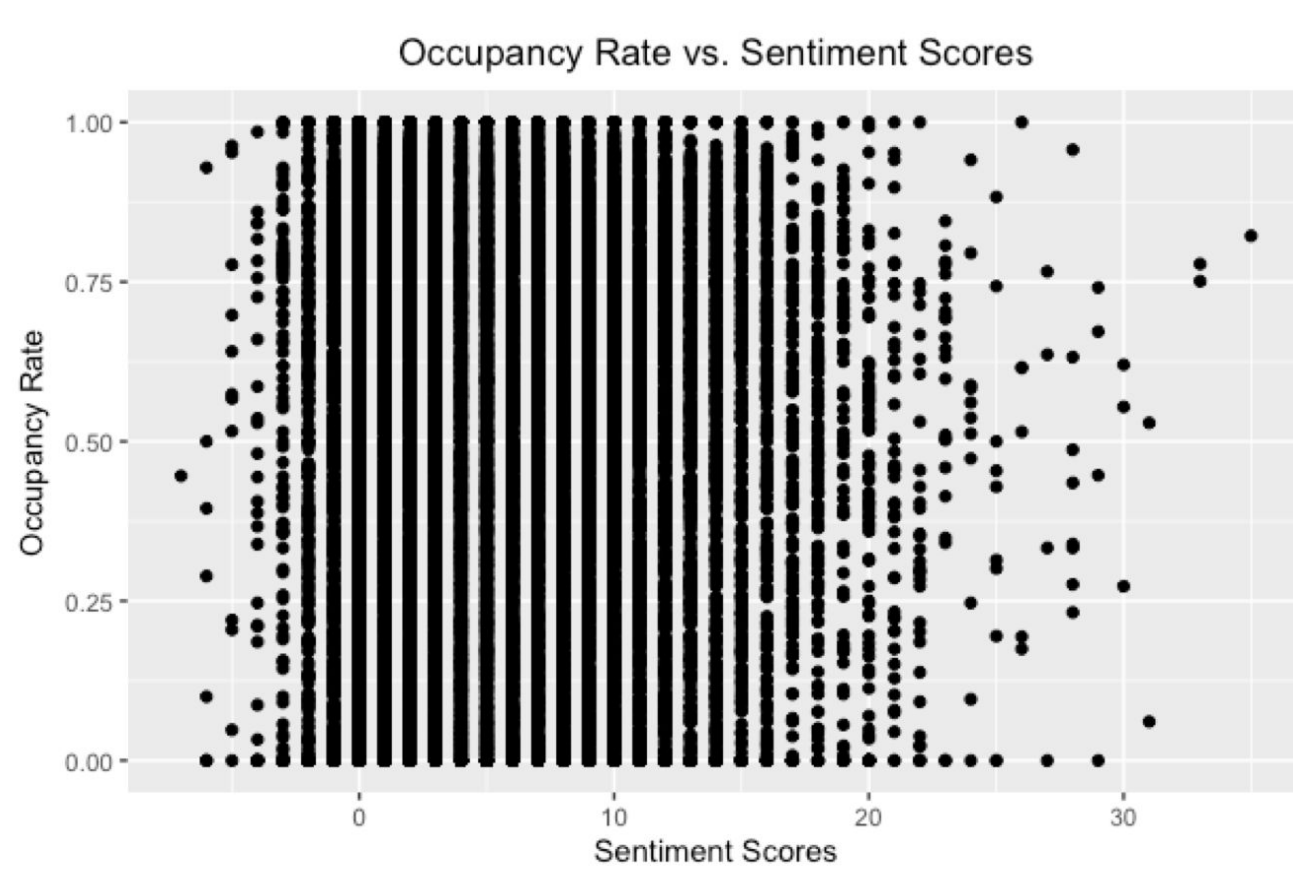

The graph shows very little indication of any correlation between occupancy rate and sentiment score. In fact, calculations reveal that this value was insignificant (at a cut-off of 0.05) with a value of only 0.048. Again, these results will certainly vary based on different methods for calculating the sentiment scores, nonetheless the low correlation does reveal that our original hypotheses may be over-exaggerated. Thus the question arises: if sentiment score does not significantly impact occupancy rate, which features do?

Part of the difficulty of working with our original data was the myriad data points that obscured visual trends. In order to better conceptualize the data, we grouped the data points into 20 distinct categories based on the level of occupancy rate. That is, all listings with an occupancy rate between 0-0.05, 0.05-0.1, (…), 0.95-1 were grouped together. Having done so, we proceeded to plot graphs of occupancy rate and sentiment scores once again:

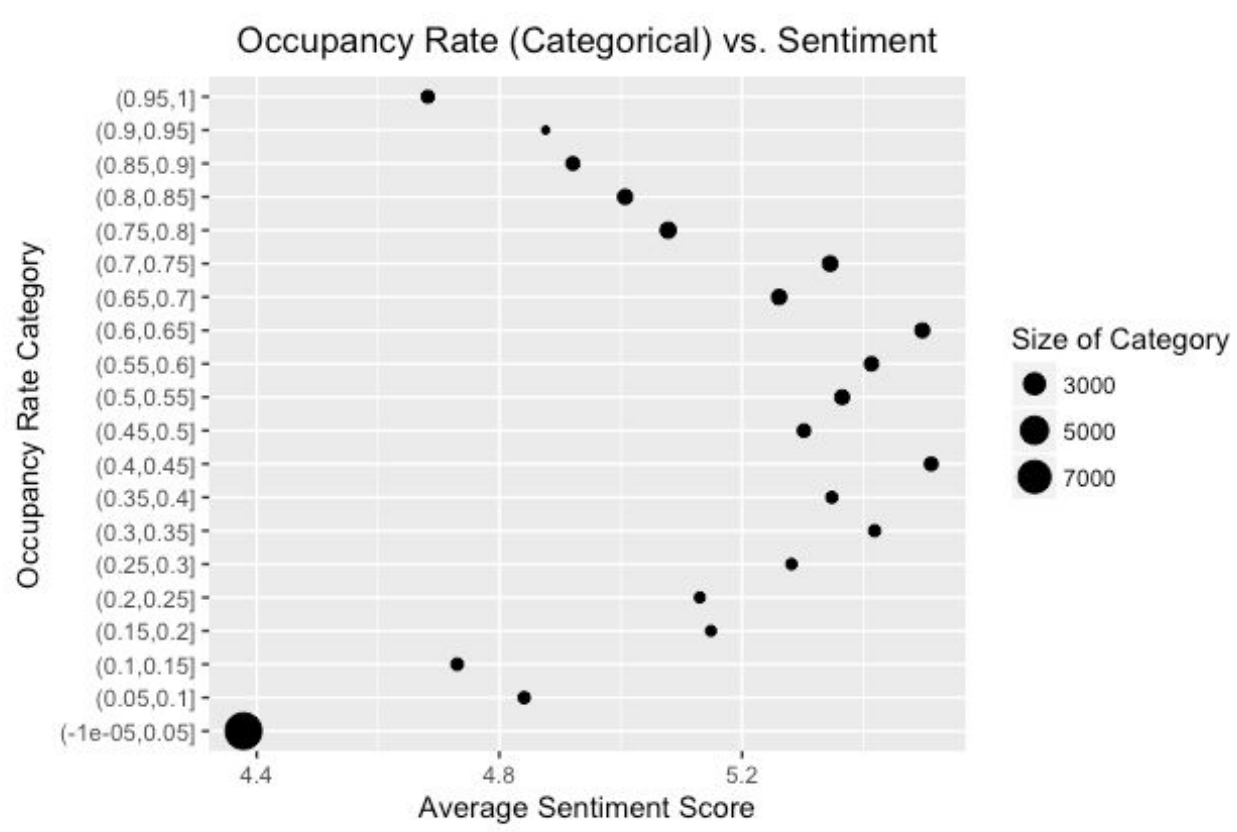

Despite the low correlation, this plot reveals the possibility that some trend exists. In particular, we observe that the average sentiment score is highest among listings within the mid-range of occupancy rates, and lowest amongst listings with either very high or very low occupancy rates. Perhaps with more advanced natural language processing techniques that can improve the accuracy of our sentiment scores, we would see this trend more pronounced. Given our particular method of calculating sentiment scores, however, we will have to find stronger correlations among other features. After creating several similar plots, we find that amenities and number of reviews are most strongly correlated with occupancy rate with correlations of (0.4 and 0.2 respectively). As expected both visualizations show noticeably stronger (positive) linear trends compared to sentiment scores.

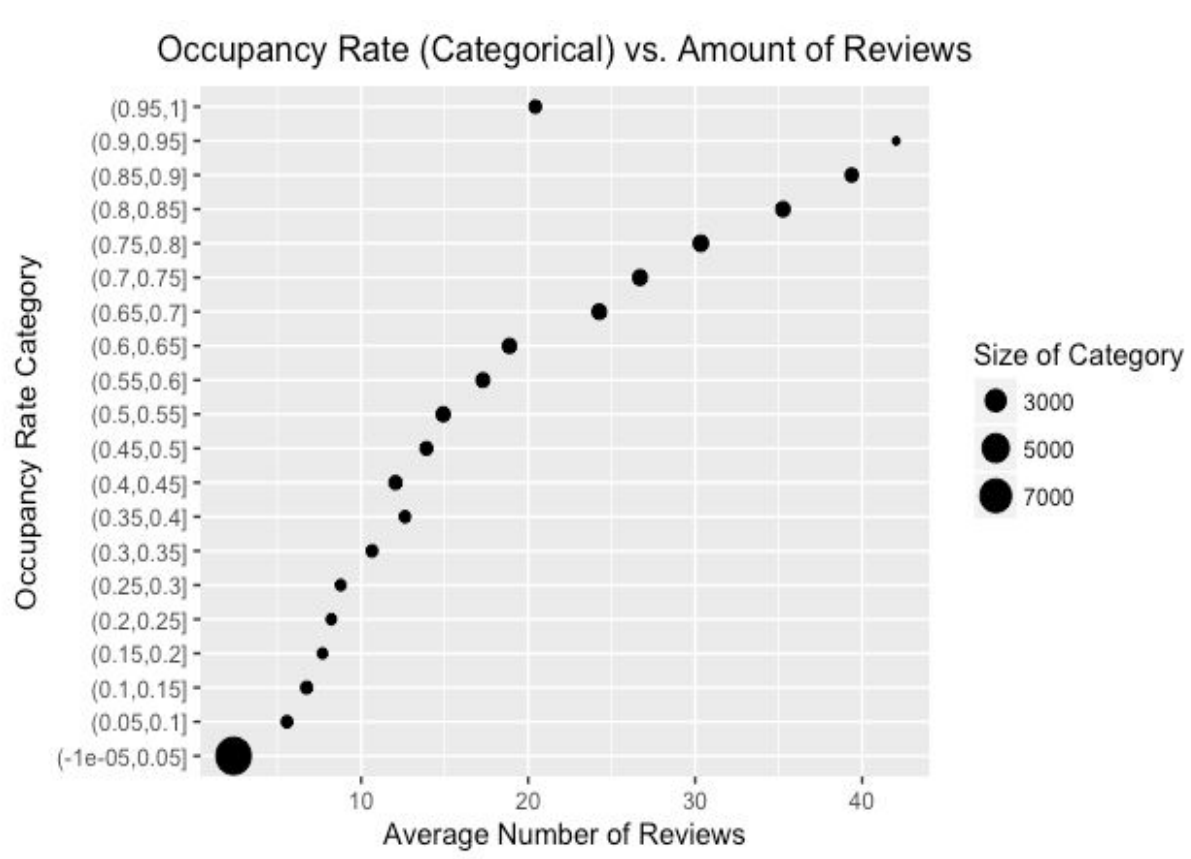

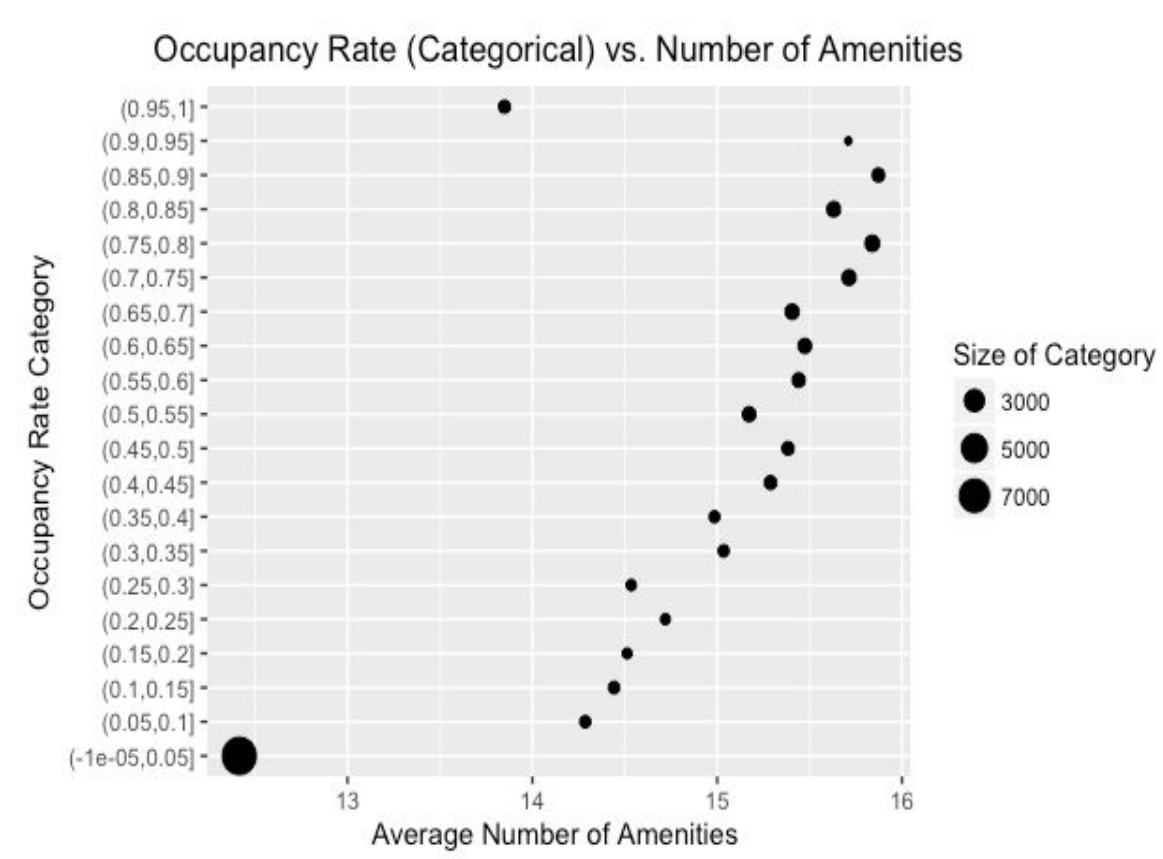

***Number of Reviews vs. Sentiment Score***

Our story as it stands is interesting: current AirBnB hosts may assume that higher sentiment scores would have a positive impact on occupancy rate. While we find that this is not the case, the fact that there is a decent correlation (R-squared of 0.4049) between number of reviews and occupancy rate means that potential correlation between number of reviews and sentiment scores could have implications regarding the causality of sentiment scores and occupancy rate. More specifically, if the number of reviews and occupancy rates are correlated, does the relationship between number of reviews and sentiment score imply something about occupancy rates? However, as shown below, we demonstrate that despite this possible inference, number of reviews and sentiment scores have a near 0 correlation, with an R-squared of -0.0784. This nevertheless sheds light on the significance of lurking or confounding variables affecting causal relationships, and that correlation indeed does not imply causation.

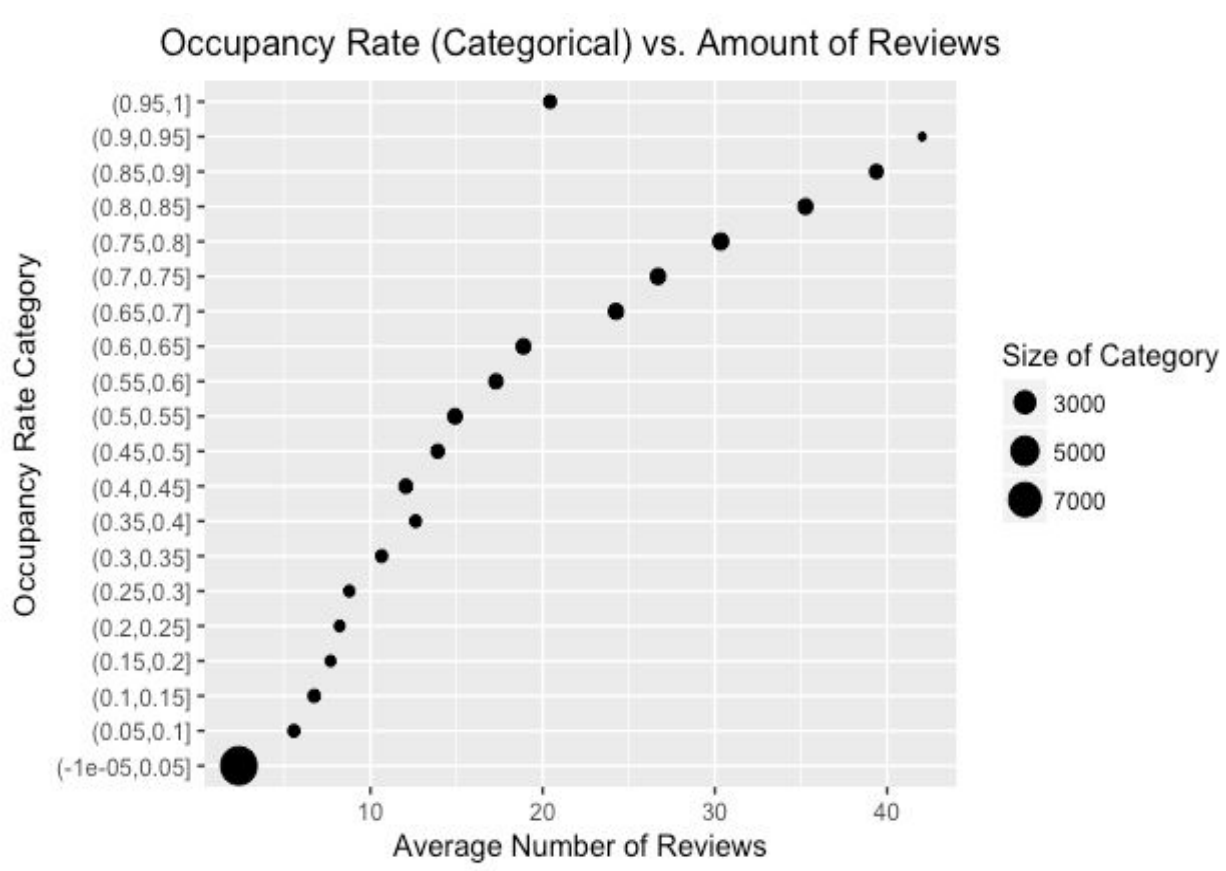

**Correlation : 0.4049**

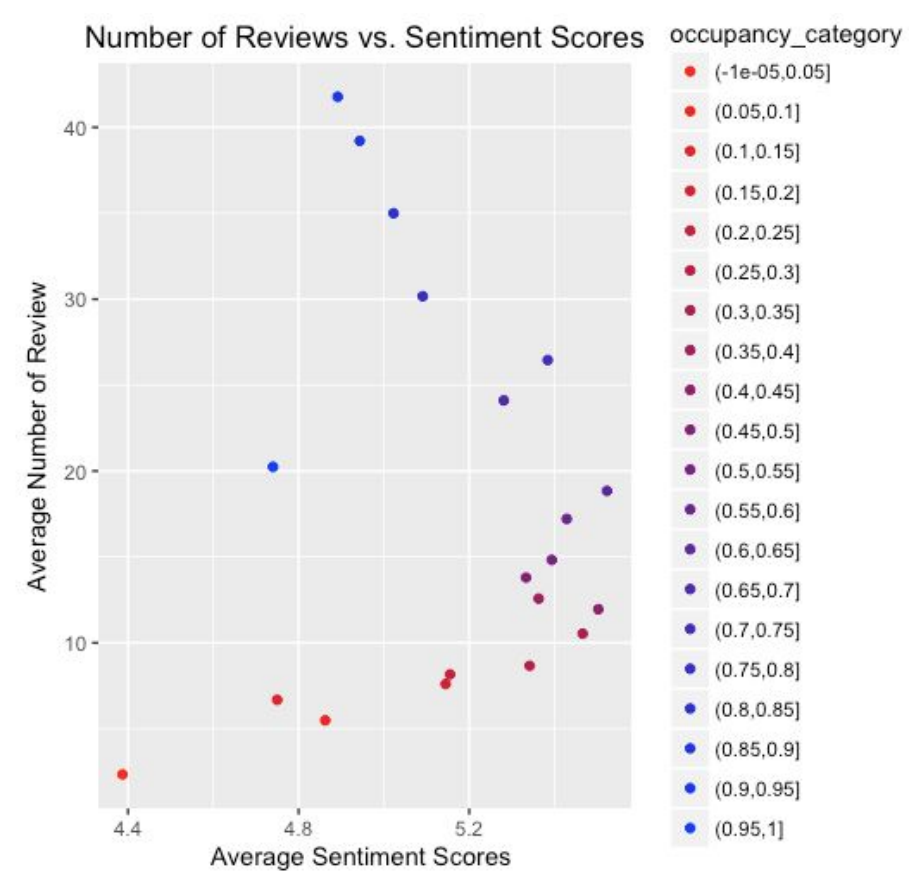

**Correlation : -0.0784**

***Forward Regression: Linear***

We decided to build a forward regression model to determine what are the most important features in determining occupancy rate, and to see how well our model is able to generate predictions. First, we created a training set by randomly selecting 80% of our listings and validation set with the remaining 20%.

Using the '*MASS*' package, which includes "Functions and datasets to support Venables and Ripley, "Modern Applied Statistics with S" (4th edition, 2002)," we used the *step* function to generate the best model using the training set. Our best model (with an AIC of *-73445.52*) was the following:

*occupancy rate ~ number of reviews + number of amenities + length of summary description*

This model has a R-squared value of *0.2086,* which means that only about 21% of the variation in 'occupancy rate' is explained by this linear model. To test how good our model is, we used the *predict* function to generate predictions of occupancy rate on the validation set. The Mean-Squared-Error of these predictions is *0.08505379*, which suggests that our model did not generate reliable predictions of occupancy rate.

We believed that there may be differences in the features that influence occupancy rate for different areas of New York City, i.e. a wealthier area like Manhattan may be different than a more affordable neighborhood. Assuming that zip code is a good determinant of divisions in the pricing of neighborhoods, we found the mean listing price for each zip code as a measure of the affordability of a neighborhood. The overall mean and standard deviation of listing prices were $137 and $104, respectively. We categorized zip codes with an average listing price of more than ½ standard deviation as "expensive neighborhoods" and zip codes with an average listing price of less than ½ standard deviation as "more affordable neighborhoods." Using this definition, we found an 'expensive' zip code, an 'average' zip code, and an 'affordable' zip code, and built forward regression models for each one. Again, we separated the listings into a training set and validation set, and generated predictions to evaluate the efficacy of the three models. Below we have shown the models found for each, along with their R-squared value and mean-squared-error of their predictions.

**'Expensive' Zip Code**: 10011
*Model: occupancy rate ~ number of reviews + number of amenities + accommodates + price + number of bedrooms*
*R-Squared: 0.2253*
*MSE of Validation Set: 0.08838848*

**'Average' Zip Code**: 11211
*Model: occupancy rate ~ number of reviews + number of amenities + price + number of people accommodated + number of bedrooms + length of summary description + rating*
*R-Squared: 0.2179*
*MSE of Validation Set: 0.08796691*

**'Affordable' Zip Code**: 11237
*Model: occupancy rate ~ number of reviews + length of summary description + number of amenities + price + number of people accommodated + number of bathrooms*
*R-Squared: 0.2367*
*MSE of Validation Set: 0.07100734*

While the models generated by controlling for zip code were different from the general model, the R-squared value was still relatively low and the mean-squared-error of the predictions were high, suggesting that these models did not produce more accurate predictions. It is interesting to note that 'number of reviews' and 'number of amenities' are still amongst the most important features for these models. The higher number of features that these models include also may be a function of overfitting.

Additionally, we wanted to generate a model for when we controlled for number of reviews, because we theorized that the most important features would change for listings with a certain range of reviews. To test out this hypothesis, we controlled for listings with between 30 and 50 reviews, calling them "medium reviewed places". Controlling for this range of reviews, we found the best model (below).

*occupancy rate ~ number of amenities + length of summary description + length of space description + rating + number of people accommodated + bedrooms*

Unsurprisingly, number of amenities and length of summary description are still among the most important features. There are additional features in this model, including length of space description, rating, number of people accommodated, bedrooms. However, the low R-squared value of 0.08575 and relatively high MSE of 0.05579298 makes this model an unreliable predictor of occupancy rate.

***Forward Regression: Logistic***

Since our linear models did not generate good predictions of occupancy rate, we proceeded to try a logistic model to predict occupancy rate in categories. To do this, we placed occupancy rates in categories of 10%- from 0.1-0.2, 0.2-0.3, and so forth. We used the same 'step' function to do a forward regression on multi-nominal logistic models, and found the best model, shown below:

*occupancy category ~ number of reviews + length of summary description + number of amenities + number of beds + price + number of people accommodated + length of space description + rating + sentiment of summary description*

Similar to the linear model, the 'number of reviews', 'length of summary description' and 'number of amenities' were amongst the most important features. The accuracy of the predictions generated by this logistic model is 0.3267658, which is better than predicting occupancy category based on frequencies.

***What about Price and Occupancy?***

Since pricing of a residence is an important feature to hosts, and because of our general objective to help hosts select the best features to improve occupancy rate, we dedicate this next section to the relationship between total price and occupancy rate.

Interestingly enough, we found no clear relationship between the total price of a listing and occupancy rate. As elucidated in the previous section, although total price was included as one of the features in our data set, it did not appear as a significant feature in the best forward-regression built occupancy rate model. More specifically, the correlation between price and occupancy rate is -.0245 – which is weak, and as intuitively makes sense, negative.

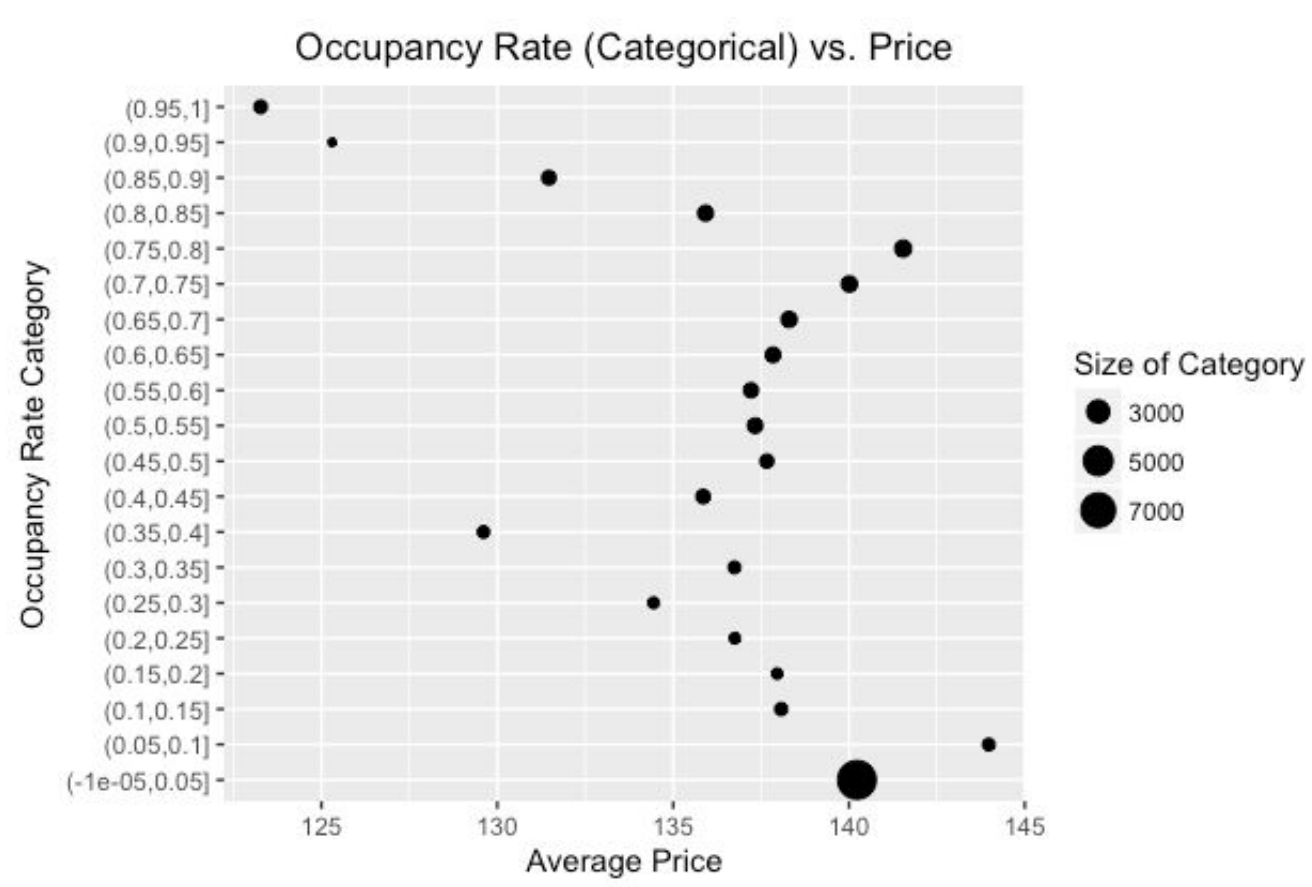

The occupancy rate (categorical) vs. price visualization above depicts the average total price of listings that correspond with specific categorized occupancy rates. One will notice that, generally, for each occupancy rate category, the average residence price falls between the same range of $135 - $140 (with most having zero occupancy). The average price of a residence in the entire New York City data set is $137, and thus, the fact that most of the dots in the graph fall in that price range shows how uncorrelated the variables are. Much of the weakness in correlation is due to the fact that occupancy rate, besides at zero, is very evenly distributed across the categories, regardless of price changes. Even so, it is

interesting to note outliers, such as the very high occupancy rates with average listings, price less than or around $125 (extreme far upper left dots). These outliers could indicate price thresholds that heavily decrease or increase occupancy rates, despite the fact that the two variables are uncorrelated over their overall ranges. Although we were limited by the data to make meaningful observations on these smaller price and occupancy ranges, we hope to be able to pursue this as further research.

Our original hope, if we had found a sizeable correlation between price and occupancy rate, was to be able to recommend prices that a host should set for targeted increases in host occupancy rate. We would accomplish this using a regression model that made price the dependent variable and occupancy rate a feature. Unfortunately, as expected, when we made this new forward stepwise regression model, we found the best price model included only the *number of people a host can accommodate* as a feature influencing occupancy rate, which naturally makes sense but is largely unhelpful. This further cemented the weak relationship between total listing price and occupancy rate.

Pivoting from the *total price* of a listing, we decided it was pertinent to examine a new feature that we had not yet looked at: *price-per-occupant* (price / accommodates). This feature, instead of the former, would perhaps be a better one to look at because it is more balanced and may be more relevant to multiple people who are looking to book and base their booking on how much they have to pay each. Indeed, the correlation between occupancy rate and price-per-occupant is -0.146, which is stronger than the correlation of -.0245 between total listing price and occupancy rate noted earlier. Moreover, when we added this feature to our starting data set and re-created the original forward regression model using all of the features in the data set, price-per-occupant was added as a feature in the best model:

*occupancy_rate ~ number_of_reviews + num_amenities + summary_length + price_per_occupant*

This new forward regression model performed slightly better than the previous without price-per-occupant, with a MSE on test data of .0839. This is marginally lower than the .085 of the first model.

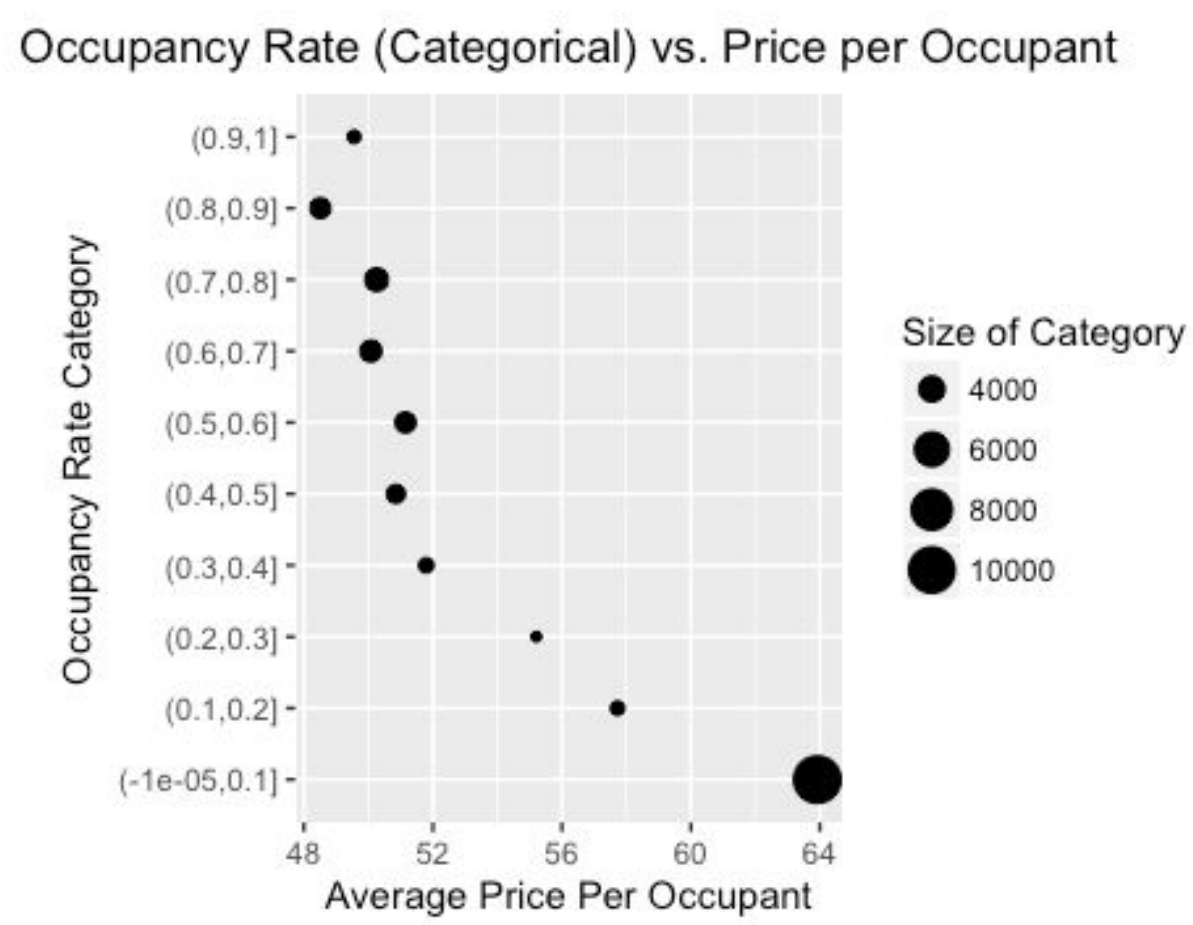

From the visualization above, one can see that negative relationship between price per occupant and occupancy rate is stronger, with lower occupancy rates having a higher average price per occupant. Because of such a relationship, we decided to test it further using the same method as before, this time building a forward regression model with price-per-occupant as the dependent variable and occupancy rate a feature. Unfortunately, much like before, the best price model was not helpful in our goal of recommending prices that a host should set. The new best price-per-occupant model only included the intercept as the best feature, which makes sense since the original total price model only included the number of accommodates (which price-per-occupant accounts for).

Overall, given our data, we were surprised not to find a strong relationship between price and occupancy rate for hosts in NYC. While there exists a weak correlation, our limited data and the distributions of prices and occupancy rates within it illustrates that the effect is minimal. While further exploration is needed, especially of outliers, thresholds, and distribution patterns – this weak correlation information is useful in and of itself as it allows hosts to worry less about how to price of listings.

# IV. Limitations of Our Analysis

***Analysis Limitations***

While we initially intended to analyze a causal relationship, our project crystallized the limitations of doing so due to unconfoundedness assumptions. Had we found a statistically significant correlation between sentiment score and occupancy rate, our goal was to design our study as close as possible to a natural experiment design (in the case that a RCT was difficult to pursue) as a valuable indicator of causal inference. In such a case, we would attempt to define our treatment and control while isolating for previously mentioned confounding variables to infer how host sentiment impacts occupancy rate. In order for this causal relationship to be valid, the stable unit treatment value assumption (SUTVA) assuming non-interference and no variation must have been satisfied, in addition to unconfoundedness assumptions. No variation would have been satisfied since all hosts were evaluated by the same sentiment score analysis we performed in our study. Non-interference seems to be satisfied, as any host with a high sentiment score should not directly affect the occupancy rate of other hosts. Additionally, there are many other reasons for low occupancy rates which are not influenced by another host's listing description, or the features we have included. Unconfoundedness, however, would be difficult to control for. Even with randomization and grouping by price, listing size, apartment type, zip code, it would have to be assumed that other confounding variables such as a host's race, age, gender, have already been controlled for.

***Data Collection Limitations***

We attempted to control and group by as many features as possible, including borough (zip code), review, seasonal variation (time of year), property type, and price of listing. While we ultimately found that sentiment score does not influence occupancy rates, we found that other features such as number of reviews, number of amenities, and summary length *could* be better indicators of a host's occupancy rate. However, we recognize that we were limited by features in AirDNA's dataset, and that there are many other features that were not included but could have impacted a host's occupancy rate. We were also subject to potential selection bias as we artificially selected feature columns to include.

Our study also shed light on the trade-off between omitted variable bias (OVB) and multicollinearity, since variables such as price and number of bedrooms (both of which we tested as features in our model) are highly correlated with each other. We attempted to mitigate this multicollinearity issue by performing a ridge regression (lasso), and carefully selecting which feature columns we believed were most significant. Nevertheless, we recognized the difficulty of striking a balance between OVB (by omitting significant variables) and avoiding the risk of multicollinearity (by including *too* many variables) and potentially overfitting.

***External Validity Limitations & Future Avenues of Research***

Despite assuming that NYC would be an accurate representation of U.S. AirBnB data, we were still limited in scope due to the local average treatment effect potentially violating external validity, and recognize that our results may not be generalizable to a larger population. If hosts from AirBnB were chosen at random from different locations, then this could provide higher external validity. However, a lack of external validity does not deem our recommendations less significant. Natural experiments are one of the few methods for strong causal inference, and we could perhaps run a natural experiment in another country. Extending our study design to other locations and performing a comparison of NYC to another city is a further avenue of research for our study. It is also important to note that a positive or negative estimated ATE does not provide information regarding how individual host's booking occupancy rate may be impacted by their sentiment score or listing description length, since we estimate the *average* treatment effect (not the individual treatment effect). Our study provided examples which crystallized this: Mary had an extremely high sentiment score of 35 and Jeff had an extremely low sentiment score of -7. While our model would suggest an insignificant difference in occupancy rate between the two hosts, we find that Mary has statistically significant higher occupancy rate of 55.4% (0.554) compared to Jeff, who had an occupancy rate of 20.5% (0.205).

## V. Conclusion

Our objective was to determine whether the way an AirBnB listing description is written causally affects occupancy (booking) rates. Our motivation for our research developed after reading multiple previous literature (described in Literature Review section) analyzing the way product review sentiment impacted sales, and was an indicator for brand loyalty. We were curious to determine whether linguistic features such as sentiment and length of a message could subjectively impact a tenant's decision to book a particular listing. Our decision to select a major city such as New York City (NYC) was founded in our desire to make our study generalizable and externally valid. After several attempts to improve our best model through a linear regression, a logistic regression, and adding quadratic terms to bring the tail of the model concave slightly, there were no significant changes to MSE. Ultimately, we found that more positive listing description did not lead to higher occupancy rates.

Pivoting from occupancy rates, we examined whether we could recommend a price for an AirBnB host to set when their occupancy rate increases or decreases, by varying other features in our model. After running another forward step regression for price, the best model indicated the only feature to be number of people a host can accommodate, a feature that the host is limited by due to size of their listing (which they cannot control). We attempted to assess how price and occupancy rate are correlated, assuming that people with higher sentiment scores may command a higher price. However, from our analysis, we have shown that more positive scores do not actually lead to higher occupancy rates. While AirBnB hosts may initially think that more positive reviews will allow them to raise their price and thus have a higher occupancy rate, we can use our analysis to show AirBnB hosts that is not the case.

Despite the discussed limitations of our analysis, our study has significant implications regarding recommendations for AirBnB hosts to reach their target occupancy rate. While AirBnB hosts may initially assume that having more positive descriptions will increase their occupancy rate as we hypothesized, we can leverage our analysis to prove this is not true. We can inform them from our analysis that price adjustments also will not significantly impact their occupancy rate. Rather, from our average summary length vs. occupancy rate scatterplot, we can recommend to hosts that there seems to be a 'good' average summary length threshold of ~ 44 words that they should incorporate to *potentially* increase their occupancy rate. We can use our results to consult AirBnB hosts with more effective ways to increase their occupancy rates by adjusting features from our best model – such as number of reviews, number of amenities, and summary length – instead of the perceived sentiment of their descriptions as we had initially assumed.

# References

Chevalier, Judith, and Dina Mayzlin. "The Effect of Word of Mouth on Sales: Online Book Reviews." *JournalofMarketingResearch*(2006).Web.<https://msbfile03.usc.edu/digitalmeasures/mayzlin/intellcont/chevalier_mayzlin06-1.pdf>.

Cohen, Levin "Empirical studies of innovation and market structure" Handbook of industrial organization, 1989.

Does the Internet Make Markets More Competitive? Evidence from the Life Insurance Industry, by Jeffrey R. Brown and Austan Goolsbee in The Journal of Political Economy, Vol. 110, No. 3 (Jun., 2002), pp. 481-507.

Decker, Reinhold, and Trusov, Michael. "Estimating aggregate consumer preferences from online product reviews". *International Journal of Research in Marketing* 27.4 (2010): 293-307.

Liu, Bing. *Sentiment analysis and Opinion Mining*. San Rafael, CA: Morgan & Claypool, 2012. 2012. Web. <http://www.morganclaypool.com/doi/pdf/10.2200/S00416ED1V01Y201204HLT016>.

McElheran, Kristina Stefansson. "The Effect of Market Leadership in Business Process Innovation: the Case of E-Business Adoption," Management Science; Issue: 6; 2015; Pages: 1197-1216.

Suryadi, Dedy, and Harrison Kim. "Identifying the Relations Between Product Features and Sales Rank From Online Reviews." *Volume 2A: 42nd Design Automation Conference* (2016). Web.

# Appendix

*Appendix 1: Original AirBnB Dataset*

| | xl_picture_url | host_id | host_url | host_name | host_since | host_location | host_about | host_response_time | host_response_rate |
|---|---|---|---|---|---|---|---|---|---|
| .. | https://a1.muscache.com/im/picture... | 3493067 | https://www.airbnb.com/users/show... | Matt | 2012-09-06 | New York, New York, United States | I'm a musician and music producer c... | within a few hours | 100% |
| .. | https://a2.muscache.com/im/picture... | 49027539 | https://www.airbnb.com/users/show... | Rob | 2015-11-14 | New York, New York, United States | Just me, simple living, clean life, hap... | within a day | 60% |
| .. | https://a0.muscache.com/im/picture... | 4712698 | https://www.airbnb.com/users/show... | Jonathan | 2013-01-15 | New York, New York, United States | My name's Jonathan and I live to trav... | within a day | 100% |
| .. | https://a2.muscache.com/im/picture... | 21492250 | https://www.airbnb.com/users/show... | Zhenyu | 2014-09-18 | New York, New York, United States | NA | N/A | N/A |
| .. | NA | 5460788 | https://www.airbnb.com/users/show... | Ken | 2013-03-14 | New York, New York, United States | NA | within a few hours | 100% |
| .. | NA | 31913440 | https://www.airbnb.com/users/show... | Courtney | 2015-04-25 | New York, New York, United States | NA | N/A | N/A |
| .. | https://a0.muscache.com/im/picture... | 4260966 | https://www.airbnb.com/users/show... | Dan | 2012-11-27 | New York, New York, United States | 37 y/o, professional, male, NYC native | within an hour | 100% |
| .. | NA | 37241813 | https://www.airbnb.com/users/show... | Samuel | 2015-07-01 | US | NA | N/A | N/A |
| .. | https://a2.muscache.com/im/picture... | 33889947 | https://www.airbnb.com/users/show... | Gautier | 2015-05-21 | New York, New York, United States | French aerospace consultant | within a few hours | 100% |
| .. | https://a0.muscache.com/im/picture... | 4158981 | https://www.airbnb.com/users/show... | Louise | 2012-11-15 | New York, New York, United States | I am a working professional in the pe... | within an hour | 100% |
| .. | https://a2.muscache.com/im/picture... | 48423154 | https://www.airbnb.com/users/show... | Evelyn | 2015-11-07 | New York, New York, United States | NA | N/A | N/A |
| .. | https://a2.muscache.com/im/picture... | 2214495 | https://www.airbnb.com/users/show... | Hayeon | 2012-04-24 | New York, New York, United States | -From New York -Love trying new foo... | within a day | 100% |
| .. | https://a2.muscache.com/im/picture... | 7812791 | https://www.airbnb.com/users/show... | Meghan & Nate | 2013-07-29 | New York, New York, United States | We are a fun, normal couple living th... | N/A | N/A |
| .. | https://a2.muscache.com/im/picture... | 1997269 | https://www.airbnb.com/users/show... | Larry | 2012-03-24 | New York, New York, United States | I'm a lover of art, beauty, language, c... | within a few hours | 90% |
| .. | https://a2.muscache.com/im/picture... | 801883 | https://www.airbnb.com/users/show... | CZ Casa | 2011-07-10 | New York, New York, United States | I've lived in NYC for almost 20 years. ... | within an hour | 100% |
| .. | https://a2.muscache.com/im/picture... | 105359406 | https://www.airbnb.com/users/show... | Lynne | 2016-11-27 | US | NA | N/A | N/A |
| .. | https://a0.muscache.com/im/picture... | 6573741 | https://www.airbnb.com/users/show... | Kim | 2013-05-25 | US | NA | within an hour | 100% |
| .. | https://a2.muscache.com/im/picture... | 49027658 | https://www.airbnb.com/users/show... | Loretta | 2015-11-14 | US | NA | N/A | N/A |
| .. | https://a2.muscache.com/im/picture... | 44379316 | https://www.airbnb.com/users/show... | Misa | 2015-09-16 | US | NA | within a few hours | 100% |
| .. | https://a2.muscache.com/im/picture... | 2510744 | https://www.airbnb.com/users/show... | Isabelle | 2012-05-31 | New York, New York, United States | I am a French-American designer and... | within an hour | 100% |
| .. | NA | 21722916 | https://www.airbnb.com/users/show... | Shailly | 2014-09-24 | New York, New York, United States | NA | N/A | N/A |
| .. | https://a1.muscache.com/im/picture... | 38583557 | https://www.airbnb.com/users/show... | Ian | 2015-07-15 | New York, New York, United States | NA | N/A | N/A |
| .. | https://a2.muscache.com/im/picture... | 59514978 | https://www.airbnb.com/users/show... | Sam | 2016-02-19 | AU | NA | within a day | 50% |
| .. | https://a2.muscache.com/im/picture... | 17090214 | https://www.airbnb.com/users/show... | Pamela | 2014-06-22 | New York, New York, United States | I'm a native NYer in the Fashion & Tr... | within a day | 100% |
| .. | https://a1.muscache.com/im/picture... | 45670064 | https://www.airbnb.com/users/show... | Antonio | 2015-10-03 | New York, New York, United States | Im 29 yrs. old and moved here a dec... | N/A | N/A |
| .. | https://a1.muscache.com/im/picture... | 5479559 | https://www.airbnb.com/users/show... | Michael | 2013-03-15 | New York, New York, United States | Hi. I'm a single parent of two adorabl... | within a day | 91% |
| .. | https://a1.muscache.com/im/picture... | 4274268 | https://www.airbnb.com/users/show... | Konstantino | 2012-11-28 | New York, New York, United States | In from Brooklyn NY | N/A | N/A |
| .. | https://a0.muscache.com/im/picture... | 14892152 | https://www.airbnb.com/users/show... | Laura | 2014-04-29 | New York, New York, United States | We have lived and worked abroad ou... | N/A | N/A |

*Appendix 2: Data Cleaning*

| host_id | name | summary | space | property_type | accommodates | bathrooms | bedrooms | beds | bed_type | price_in_dollars | number_of_reviews | zipcode | occupancy_rate |
|---|---|---|---|---|---|---|---|---|---|---|---|---|---|
| 3493067 | best studio on prospect park | perfect studio for couples or individu... | the space: the apartment is a good si... | Apartment | 2 | 1.0 | 1 | 1 | Futon | 75 | 145 | 11225 | 0.895 |
| column 1: numeric with range 43 - 105988052 | | well located 1 bedroom apt on t... | the apartment is yours for your stay; ... | Townhouse | 5 | 1.0 | 1 | 2 | Real Bed | 117 | 50 | 11211 | 0.932 |
| 4712698 | private room, 10min to times square | this 2br apartment in lic offers a priv... | update: my apartment was recently f... | Apartment | 2 | 1.0 | 1 | 1 | Real Bed | 89 | 43 | 11101 | 0.665 |
| 21492250 | clean and tidy separated livingroom | it is the living room separated in an ... | NA | House | 2 | 1.0 | 1 | 1 | Real Bed | 80 | 0 | 10025 | 0.000 |
| 5460788 | awesome one bedroom gramercy park | dear guest, welcome to new york! tha... | [details] i'll let the pictures speak for ... | Apartment | 4 | 1.0 | 1 | 1 | Real Bed | 200 | 1 | 10003 | 0.385 |
| 4260966 | beautiful, central,sunny, quiet shared... | this apartment is located in the heart... | clean and modern, with a renovated ... | Apartment | 1 | 1.0 | 1 | 1 | Real Bed | 119 | 1 | 10009 | 0.939 |
| 33889947 | beautiful one bedroom in soho | the apartment is on king street, close... | NA | Apartment | 2 | 1.0 | 1 | 1 | Real Bed | 200 | 14 | 10014 | 0.834 |
| 4158981 | spacious uws apt. by central park | enjoy your stay in a spacious, beautif... | come experience the rhythm and rha... | Apartment | 2 | 1.5 | 1 | 1 | Real Bed | 125 | 96 | 10023 | 0.894 |
| 48423154 | brand new 1 br apt steps from train | brand new building with modern dec... | NA | Apartment | 4 | 1.0 | 1 | 2 | Real Bed | 85 | 0 | 11235 | 0.400 |
| 2214495 | spacious studio near prospect park! | spacious, private studio available - a/... | NA | Apartment | 1 | 1.0 | 1 | 1 | Real Bed | 80 | 0 | 11225 | 0.450 |
| 7812791 | park views! upper manhattan 1br | NA | available june 24, 2015 through july ... | Apartment | 2 | 1.0 | 1 | 1 | Real Bed | 79 | 0 | 10040 | 0.000 |
| 1997269 | the best of union square & chelsea | centrally located between union squa... | this beautiful modern 725 sq ft. pre-... | Apartment | 2 | 1.0 | 1 | 2 | Real Bed | 245 | 19 | 10011 | 0.811 |
| 801883 | nice clean style apt. central midtwn n... | my place is near highland park, javits... | the living room have a beautiful desi... | Apartment | 4 | 1.0 | 1 | 1 | Real Bed | 110 | 4 | 10001 | 0.688 |
| 105359406 | spacious 1-bd, sleeps 4, 20 min to ti... | full access to large 1-bedroom apart... | NA | Apartment | 4 | 1.0 | 1 | 2 | Real Bed | 150 | 0 | 11104 | 0.588 |
| 6573741 | gorgeous west village 1-bdm apt | this gorgeous 1-bedroom apt is locat... | NA | Apartment | 2 | 1.0 | 1 | 1 | Real Bed | 250 | 19 | 10014 | 0.881 |
| 49027658 | apartment 20 minutes from manhattan | quick access to midtown manhattan (... | NA | Apartment | 4 | 1.0 | 1 | 2 | Real Bed | 143 | 0 | 11222 | 0.000 |
| 44379316 | spacious private room in brooklyn | located in a lively neighbourhood in ... | NA | Apartment | 1 | 1.0 | 1 | 1 | Real Bed | 40 | 0 | 11237 | 0.450 |
| 2510744 | large private 1br with backyard - willi... | quiet, comfortable and 100% private ... | beautiful newly renovated 1 bedroom... | Apartment | 2 | 1.0 | 1 | 1 | Real Bed | 98 | 6 | 11211 | 0.000 |
| 21722916 | lovely nyc apartment july 16-30 | a light filled apartment on the 20th fl... | NA | Apartment | 4 | 1.0 | 2 | 2 | Real Bed | 400 | 0 | 10028 | 0.133 |
| 38583557 | bohéme 1br steps from prospect park | beautiful, spacious 1 br with 5 minut... | this place was a labor of love! at this ... | Apartment | 4 | 1.0 | 1 | 2 | Real Bed | 99 | 2 | 11226 | 0.000 |
| 59514978 | luxury 1 br in luxury doorman building | my place is close to brooklyn bridge, ... | the apartment is located in a luxury ... | Apartment | 2 | 1.0 | 1 | 1 | Real Bed | 190 | 0 | 10006 | 0.030 |
| 17090214 | highrise studio-walk 2 times square | share my cozy midtown west space i... | this is a spacious studio apartment i... | Apartment | 1 | 1.0 | 1 | 1 | Pull-out Sofa | 100 | 9 | 10019 | 0.044 |
| 45670064 | private room great area | great neighborhood. 2 blocks from 6... | the room is open and comfortable. q... | Apartment | 2 | 1.0 | 1 | 1 | Real Bed | 60 | 2 | 10029 | 1.000 |
| 5479559 | bedroom for two in chelsea | NA | clean, safe, and, most of all, inexpen... | Apartment | 2 | 1.0 | 1 | 1 | Real Bed | 80 | 109 | 10011 | 0.945 |
| 4274268 | apt in prospect/crown heights | 2,3,4,5 - s shuttle, b,q subway access... | this apartment is a bright, one bedro... | Apartment | 4 | 1.0 | 1 | 2 | Real Bed | 110 | 3 | 11216 | 1.000 |

*Appendix 3: Frequency of Phone Features in Literature Review (Decker and Trusov 2010)*

Table 3.
Extracted functional attributes and associated relative frequencies [in %].

| Attribute | Freq. | Attribute | Freq. | Attribute | Freq. |
|---|---|---|---|---|---|
| Size/weight | 33.5 (34.8) | Menu navigation | 8.9 (10.0) | Interfaces | 4.1 (4.4) |
| Appearance ("look") | 23.2 (23.5) | Camera | 7.8 (10.2) | Robustness/stability | 3.9 (5.8) |
| Equipment/functionality[a] | 21.7 (22.9) | Memory | 7.2 (8.4) | Manufacturing quality | 3.7 (7.6) |
| Battery | 16.8 (20.1) | Other functions | 6.7 (7.8) | Software | 3.2 (6.3) |
| Price | 16.0 (17.1) | Mobile Internet | 6.7 (8.2) | Radio technology | 3.0 (3.4) |
| Operation/handling | 15.5 (18.4) | Keypad/touch screen | 6.2 (7.3) | Sound | 2.5 (3.3) |
| Display | 12.6 (14.2) | Messaging | 5.8 (3.9) | Reliability | 1.0 (1.9) |
| Multimedia | 9.0 (11.0) | Reception/voice quality | 5.8 (8.3) | | |

a Functionality is referred to in the meaning of "variety of functions."

*Appendix 4: Sentiment Score Methodology*

```
# -------- Calculating Score -------- #

description_m <- matrix(nrow= length(listing_descriptions),
                        ncol = length(sentiment_scores$score))
description_m[] <- 0L

# for each of the words check if it is in the vector
# if it is in the vector, get the index of the vector
# and then update the according element [i,j] in the matrix

for(i in 1:length(listing_descriptions)){
  current_listing = listing_descriptions[i]
  words_listing <- strsplit(paste(current_listing, collapse = " "), ' ')[[1]]
  words_listing <- words_listing[words_listing != ""]
  for(j in 1:length(words_listing)){
    current_word = gsub("[[:punct:]]","",words_listing[j])
    #Removing exclamation marks, periods, and blank spaces
    if (current_word %in% sentiment_scores$word){
        j_index <- match(current_word,sentiment_scores$word)
        description_m[i,j_index] <- description_m[i,j_index] + 1
    }
  }
}

final_scores <- description_m[] %*% sentiment_scores$score
listings_clean <- mutate(listings_clean, sentiment_scores = final_scores)
```